# From anti-Arrhenius to Arrhenius behavior in a dislocation-obstacle bypass: Atomistic Simulations and Theoretical Investigation

Mohammadhossein Nahavandian[a], Soumit Sarkar[a], Soumendu Bagchi[b], Danny Perez[b], Enrique Martinez[a,c]

[a] *Department of Mechanical Engineering, Clemson University, Clemson, SC 29623, USA*
[b] *Theoretical Division, Los Alamos National Laboratory, Los Alamos, New Mexico 87544, USA*
[c] *Department of Materials Science and Engineering, Clemson University, Clemson, SC 29623, USA*

**Abstract**

Dislocations are the primary carriers of plasticity in metallic materials. Understanding the basic mechanisms for dislocation movement is paramount to predicting the material mechanical response. Relying on atomistic simulations, we observe a transition from non-Arrhenius to Arrhenius behavior in the rate of an edge dislocation overcoming the long-range elastic interaction with a prismatic loop in tungsten. Close to the critical resolved shear stress, the process shows a non-Arrhenius behavior at low temperatures. However, as the temperature increases, the activation entropy starts to dominate, leading to a traditional Arrhenius-like behavior. We have computed the activation entropy analytically along the minimum energy path following Schoeck's method [1], which captures the cross-over between anti-Arrhenius and Arrhenius domains. Also, the Projected Average Force Integrator (PAFI) [2], another simulation method to compute free energies along an initial transition path, exhibits considerable concurrence with Schoeck's formalism. We conclude that entropic effects need to be considered to understand processes involving dislocations bypassing elastic barriers close to the critical resolved shear stress. More work needs to be performed to fully understand the discrepancies between Schoeck's and PAFI compared to molecular dynamics.

**Keywords**: Activation Entropy, Activation Volume, Molecular Dynamics Simulation, Dislocation Dynamics, Arrhenius/Anti-Arrhenius

# 1. Introduction

It is well known that dislocations within a crystalline material play a crucial role in the long-term plastic deformation. Thermally activated dislocation processes have been extensively studied in the literature [3–6]. The related free energy barrier for dislocation glide directly affects the time a dislocation will need to start moving. Thermal lattice vibrations and the stresses that cause the dislocation to move affect the rate for the dislocation to overcome free energy barriers posed by different types of obstacles [7]. By concentrating on the minimum energy pathway (MEP), transition state theory offers a formalism to estimate the rate of such processes [8,9]. Numerous phenomena in materials science, such as metallic alloy plasticity [10], or creep [11], depend on thermally activated mechanisms. Transition state theory [12] relates the activation free energy $\Delta G$ with the rate for the process at constant stress:

$$\Gamma_{TST} = \nu \exp\left(-\frac{\Delta G}{k_B T}\right) \qquad (1)$$

where $\nu$ is a pre-exponential factor, $k_B$ Boltzmann constant, and $T$ the temperature. According to this expression, we need to compute both $\Delta G$ and $\nu$ to predict the rate at a given temperature with $\Delta G = \Delta E_0 + \Delta E_{el} + \Delta E_p - T\Delta S$, with anti-Arrhenius behavior occurring when the free energy barriers depend on temperature and the dependence is stronger than linear [13]. Atomistic methods can be used to compute the rate by running dynamic simulations to calculate the average time for the process, which is the inverse of the rate. Each term in the expression for $\Delta G$ can also be computed with atomistic methods. $\Delta E_{el}$ is the elastic activation energy, $\Delta E_p$ represents the plastic activation energy, and $T\Delta S$ is the entropic activation term. $\Delta E_0$ is an activation potential energy at zero stress, that can be obtained relying on the nudged elastic band (NEB) algorithm [14,15]. The elastic activation energy expression $\Delta E_{el} = \int_V [\tilde{\tau}^s \tilde{\epsilon}^s - \tilde{\tau}^i \tilde{\epsilon}^i] dV$ is a function of stress and strain at saddle point ($\tilde{\tau}^s \tilde{\epsilon}^s$) and initial ($\tilde{\tau}^i \tilde{\epsilon}^i$) state. We can assume that the strain at the initial state is zero so the elastic activation energy can be simplified as $\tilde{\tau}^s \int_V \tilde{\epsilon}^s dV$ at constant stress which finally equals $\tilde{\tau}^s \widetilde{\Delta V}$. The plastic energy expression is a result of the permanent deformation occurring in each replica with respect to the initial replica, and it can be formulated as follows: $\Delta E_p = \tilde{\tau}^s . (\bar{b} \otimes \bar{d}) \bar{l}$ where $b$ is the burgers vector, $l$ is the dislocation line length, and $d$ is the average distance swept by the dislocation line due to the imposed stress. Additional details regarding the effect of each energy term will be presented in Figure 9 in Section (3.3). $\Delta S$ is a more difficult term to obtain as it relates to the vibrational details of the system at the saddle point and the initial state. If the potential energy landscape is harmonic at both the saddle point and initial configuration, the rate expression can be simplified according to Vineyard's work [7], and the pre-exponential factor contains the entropy as independent of temperature. In cases where harmonicity does not hold, more complex approaches need to be used. This paper describes two of those approaches, projected average force integrator (PAFI) and Schoeck's [1,2]. Theoretically, the contribution of the activation entropy for metallic diffusion was explored in 1952 by Dienes [16]. In particular, an estimate of the entropy contribution caused by changes in vibrational frequencies around the saddle point was developed. Later, Dimelfi et al. [12] analyzed the entropy of a system with an internal strain field. By applying the principles of thermoelasticity in the continuum theory, they developed an analytical expression to calculate the change in entropy in a process involving strain evolution, which is applicable to elastic solids that exhibit non-linear behavior and anisotropy [12]. Later, Schoeck derived an analytical expression for the change in entropy as a function of solid materials' internal strain or stress fields [1] based on a second-order expansion of the Helmholtz free energy considering finite strains. We will use this analytical approach to rationalize the results from atomistic simulations. Ryu et al. [17] developed an approach to predict dislocation nucleation rate based on the Becker-Döring theory and umbrella sampling [18] simulations. Their findings revealed large activation entropies caused by anharmonic effects, leading to a substantial alteration in the nucleation rate by several orders of magnitude. In one case, under uniaxial tensile strain, Perez et al. [19] utilized atomistic simulations to explore the kinetics of plastic yield around small preexisting voids in copper single crystals, which were stabilized by strong entropic effects. In another study by Swinburne and Marinica, a path-based, accurate expression for free energy differences in the solid state was presented [2]. In a recent study, explicit calculations were performed to determine the anharmonic free energies, defined as the vibrational contribution beyond the quasiharmonic approximation, and entropies of Nb, Mo, Ta, and W in pristine systems. The *ab initio* computed values were employed to validate the previously established experimental data concerning the distinct trends observed in the behavior of body-centered cubic (BCC) refractory elements in groups V and VI [20].

Several studies have been conducted regarding the relationship between temperature and the reaction rate for different processes than dislocation motion. One study examines self-interstitial atom diffusion in BCC vanadium, where below 600 K, temperature-dependent correlations cause non-Arrhenius behavior, while above 600 K, small migration barriers invalidate the Arrhenius expression [21]. Wang demonstrated that as the temperature rose, activation entropy for vacancy migration in Na β"-alumina decreased [22]. Also, Cantwell et al. [23] expressed a lack of mechanistic explanations for various forms of anti-thermal behavior. However, understanding such behavior could lead to significant advancements in fields like catalysis, nanocrystalline alloys, and high-efficiency engines [23]. The effect of non-Arrhenius behavior on grain growth in ceramics based on (K, Na)NbO$_3$, and its influence on the piezoelectric properties, was also studied [24]. In another work, non-Arrhenius grain boundary migration, referred to as anti-thermal migration, was examined in an incoherent twin grain boundary in nickel [25]. Another study focused on the hot flow behavior of nickel-based superalloys by conducting compression tests at temperatures ranging from 1000-1150°C and specific strain rates. The researchers developed a modified Arrhenius constitutive relation with an activation energy map to obtain a more precise form of the model [26]. Furthermore, an experimental observation reported a non-Arrhenius transition behavior of grain growth in strontium titanate (SrTiO$_3$) [27]. It has also been found that the proportion of rapidly growing grains might show an anti-Arrhenius trend [28]. Investigating the alloy content effect on microstructure stability in copper and its alloys, specifically Cu and Cu–Al, a distinct anti-thermal temperature dependence was observed [29].

The focus of this work is to investigate how the activation entropy influences the reaction rate of an edge dislocation overcoming a barrier posed by the long-range stress field of a sessile prismatic loop in BCC tungsten under a range of temperature and stress values. To achieve this goal, we employed three distinct approaches, namely, molecular dynamics (MD) simulations, PAFI [1,2], and the theoretical Schoeck's expression [1]. Our objective is to compare the results obtained from these methods and draw conclusions about the interplay between activation entropy and enthalpy in dislocation glide.

## 2. Method

In this study, we use MD and PAFI simulations as well as Schoeck's [1] analytical expression (Equation 2) to shed light on the thermal activation of an edge dislocation overcoming the long-range interaction with an $a\langle 100 \rangle$ rhomboidal shape prismatic loop, of area equivalent to that of a circle with internal radius of 9.65 Å in tungsten (W). The center of mass of the prismatic loop is at a distance of 50.19 Å from the edge dislocation glide plane with the sample oriented in the $x = [111]$, $y = [1\bar{1}0]$, and $z = [11\bar{2}]$. Table 1 defines the simulation box dimensions, which were also used for Schoeck's and PAFI calculations, and its schematic is shown in Figure 1. We use periodic boundary conditions in x and z and free surfaces in y. We have used Marinica et al. interatomic potential [30] in all the simulations presented in this work.

We have obtained the minimum energy path of the process relying on the NEB algorithm [14,15]. As an initial guess for the path, we have interpolated linearly between initial and final configurations. This method is already implemented in LAMMPS [31] and allows us to obtain atomic configuration along the path, which will be used in PAFI and Schoeck's formalisms to estimate the free energy and entropy, respectively. We also tried to use intermediate configurations obtained directly from the MD trajectory, but the results that we found were similar to the linear interpolation.

**Table 1**. Simulation box configuration in Å

| $X_{min}$ | $X_{max}$ | $Y_{min}$ | $Y_{max}$ | $Z_{min}$ | $Z_{max}$ |
|---|---|---|---|---|---|
| -108.13 | 108.24 | -12.62 | 197.62 | -92.26 | 92.26 |

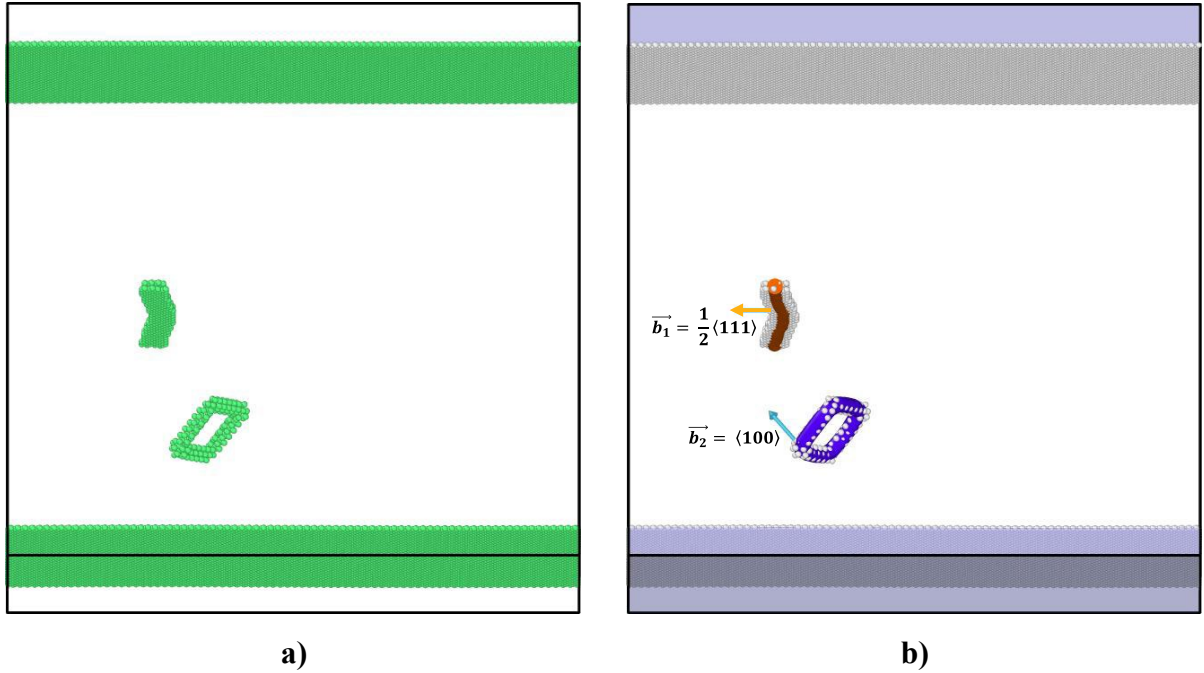

| a) | b) |

**Fig.1.** Molecular dynamics simulation **a)** 3-D atomic configuration showing atoms in a crystalline structure different from BCC according to a common neighbor analysis as implemented in Ovito [31], **b)** Dislocation line and burgers vector obtained using the dislocation extraction algorithm [32] as implemented in Ovito [33].

## 2.1. Molecular Dynamics simulation approach

Throughout this work, we have used the LAMMPS software package [31]. In order to examine the behavior of edge dislocations at different shear stress values, we conducted a comprehensive investigation over a wide temperature range spanning from 20 to 700 K. The atoms belonging to the core of the prismatic loop were frozen so not to modify the energy landscape induced by any transition involving the dislocation loop. First, the sample was thermalized for 2 ps at the target temperature using a Langevin thermostat with a damping coefficient of 1.0 ps$^{-1}$ applied in the whole system. Once thermalized, stress was applied, adding forces to the atoms in two parallel planes at both surfaces. We computed the waiting time for the dislocation to overcome the obstacle for different conditions of temperature, stress, and five independent realizations per condition. As stated above, we obtained the MEP using the NEB algorithm [34,35] at different applied stresses. To obtain the MEP at 120 MPa (above the critical resolved shear stress), the path was reconverged starting from the path at 110 MPa keeping the first replica frozen.

*2.2. Theoretical calculation of the activation entropy by Schoeck's formalism*

Schoeck developed an expression for the change in entropy induced by internal strain fields [1]. A key aspect of this derivation involves treating the solid as a continuum and utilizing the theory of elasticity for finite strains to describe the properties of defects as sources of internal strain [1]. Equation 2 presents the main result, with the change in entropy given as a function of the mechanical properties of the material and the internal strains.

$$\Delta S = \int \alpha_V K \, V_{ii} dV - \frac{1}{2} \int \frac{\partial C_{iklm}}{\partial T} V_{ik} V_{lm} dV \tag{2}$$

with $\alpha_V = 3\alpha$ where $\alpha$ and $\alpha_V$ are the linear and volumetric thermal expansion coefficients of the material, $dV$ is the volume element, $C_{iklm}$ is the 4$^{th}$-order tensor with the elastic constants and, $V_{ik}V_{lm}$, and $V_{ii}$ represent the elastic strain components of the second-order Green's strain tensor [1]. Although the tungsten ground state crystal structure is BCC and isotropic, the elastic constants deviate from the cubic symmetry in the current configuration with the free surface and the dislocations. In this case, we obtain seven independent components in the stiffness tensor, $C_{11}$, $C_{22}$, $C_{13}$, $C_{31}$, $C_{33}$, $C_{44}$ and $C_{66}$. Also, since we do not have external stress with hydrostatic component, the first term of Equation 2 would be zero [1], which we have verified computing the term. Although its contribution increases the total entropy, denoted as S, it does not modify the entropy change (ΔS) between each replica along the MEP and the initial state. Hence, we concluded that the first term can be neglected. Since we are interested in studying the temperature dependence of the free energy, we find a mathematical expression depending on both temperature and strain for the seven aforementioned elastic constants, shown below in GPa. Thus, from numerical simulations for the specific case of tungsten (see Supplementary Material) we obtain:

$$C_{11} = (0.315068 \, V_{xx}^2 - 0.038698 \, V_{xx} - 0.000317)(0.000103 \, T^3 - 0.646466 \, T^2 + 1064.446719 \, T - 1604185.571516) \tag{3}$$

$$C_{22} = 0.345T + 600.83 \tag{4}$$

$$C_{31} = (4.299493 \, V_{zz} + 0.096801)(0.000490 \, T^2 - 0.527812 \, T + 1222.310181) \tag{5}$$

$$C_{13} = 0.0334T + 109.23 \tag{6}$$

$$C_{33} = (0.073300\ V_{zz}^2 + 0.004982\ V_{zz} + 0.000147)\,(-0.001104\ T^3 \\ + 2.497100\ T^2 - 3101.287910\ T + 4270935.703701) \tag{7}$$

$$C_{44} = -0.01095T + 79.92538 \tag{8}$$

$$C_{66} = 0.0543T + 189.68068 \tag{9}$$

Upon examination of Equations (3-9), a notable observation arises when stress versus strain components are plotted. It becomes apparent that, in the majority of cases the relationship does not follow Hooke's law and exhibits a nonlinear trend (see Supplementary Material). To address this issue, we employed nonlinear regression techniques to capture both the strain and temperature dependencies of the elastic constants. Now, we can rewrite Equation 2 in discretized form as follows:

$$\Delta S = \frac{-1}{2}\left[\sum_{n=1}^{N}\frac{\partial C_{11}}{\partial T}(V_{11}V_{11})_n\,\forall_n + \sum_{n=1}^{N}\frac{\partial C_{22}}{\partial T}(V_{22}V_{22})_n\,\forall_n \right.\\
+ \sum_{n=1}^{N}\frac{\partial C_{13}}{\partial T}(V_{11}V_{33})_n\,\forall_n + \sum_{n=1}^{N}\frac{\partial C_{31}}{\partial T}(V_{33}V_{11})_n\,\forall_n \\
+ \sum_{n=1}^{N}\frac{\partial C_{33}}{\partial T}(V_{33}V_{33})_n\,\forall_n + \sum_{n=1}^{N}\frac{\partial C_{44}}{\partial T}(V_{12}V_{12})_n\,\forall_n \\
\left. + \sum_{n=1}^{N}\frac{\partial C_{44}}{\partial T}(V_{13}V_{13})_n\,\forall_n + \sum_{n=1}^{N}\frac{\partial C_{66}}{\partial T}(V_{31}V_{31})_n\,\forall_n \right] \tag{10}$$

Substituting Equations (3-9) into Equation (10), we have the final expression for entropy, which is a second order function of temperature. Also, it should be noted that the entropy values were not highly sensitive to $C_{22}$. To avoid confusion between volume and strain symbols, we denote the volume with symbol $\forall$ which represents the atomic volume in equations 10 and 11.

$$\Delta S = \frac{-1}{2}\Bigg[\sum_{n=1}^{N}[(0.315068\ V_{11}^2 - 0.038698V_{11} - 0.000317)(0.000309\ T^2$$
$$- 1.292932\ T + 1064.446719\ )(V_{11}V_{11})]_n \forall_n$$
$$+ \sum_{n=1}^{N} 0.345(V_{22}V_{22})_n \forall_n + \sum_{n=1}^{N} 0.0334(V_{11}V_{33})_n \forall_n$$
$$+ \sum_{n=1}^{N}[(4.299493V_{33} + 0.096801)(0.00098\ T$$
$$- 0.527812\ )(V_{33}V_{11})]_n \forall_n \tag{11}$$
$$+ \sum_{n=1}^{N}[(0.073300\ V_{33}^2 + 0.004982\ V_{33}$$
$$+ 0.000147)(-0.003312\ T^2 + 4.9942\ T$$
$$- 3101.287910)(V_{33}V_{33})]_n \forall_n - \sum_{n=1}^{N} 0.01095(V_{12}V_{12})_n \forall_n$$
$$- \sum_{n=1}^{N} 0.01095(V_{13}V_{13})_n \forall_n + \sum_{n=1}^{N} 0.0543(V_{31}V_{31})_n \forall_n\Bigg]$$

We computed the crystalline systems' atomic-level elastic strain and the deformation gradient tensors using OVITO [32,36]. We have performed the NEB simulations with 31 replicas along the MEP. Since the position of the saddle point is unknown, the number of replicas needs to be large enough not to skip the maximum. In principle, the saddle point in the free energy landscape might not correspond with the maximum of enthalpy, i.e., the reaction coordinate may be different. The atomic volume is computed for every atom using a Voronoi tessellation [33] excluding the atoms at the surface. This data allows us to estimate the entropy following Equation (11).

For visualization purposes, the Dislocation Extraction Algorithm (DXA) analysis was used, which effectively detects and characterizes dislocation lines within an atomistic crystal structure. DXA identifies the Burgers vectors associated with each dislocation and generates a visual representation of the dislocations in the form of line segments [32].

Once we compute the entropy using Schoeck's approach, incorporating the barrier values obtained from the NEB along the path, we can determine the Gibbs free energy and subsequently calculate the reaction rate using (TST), following Equation 1.

*2.3. PAFI method*

The PAFI algorithm is a free energy calculation method based on transition paths. It enables the computation of anharmonic free energy profiles for intricate processes without the need to define any functions for collective variables or to converge towards a minimum free energy pathway [2,37]. PAFI estimates the partition function for the atomic configurations along the minimum energy path. It constrains the forces to remain on the hyperplane of the replica perpendicular to the path, while taking into account the temperature dependence of the reaction pathway [2]. It runs dynamics in the canonical ensemble in each configuration using Langevin dynamics with the constrains mentioned above [38]. The idea to use PAFI in this work is to compute the free energy barrier and investigate the rate of reaction through TST. In PAFI simulations, we use a friction coefficient of 100 ps$^{-1}$, which implies a strong damping [39].

We use Schoeck's [1] analytical expression (Equation 11) and PAFI [2] to rationalize the results in terms of changes in the activation free energy and rate of reaction to compare with MD and understand the effect of the different terms in the expression for the free energy.

3. **Results**

First, we computed the critical resolved shear stress ($\tau_{CRSS}$) for the process, which was found to be 118 MPa. NEB simulations were subsequently performed at different shear stresses of 0, 56, 100, 110, and 120 MPa, applied to the top and bottom planes in the x [111] direction, to compute the variation of the enthalpy along the path with fitted splines in Figure 2. The change in entropy along the MEP was computed for 31 replicas as mentioned in section 2.2. The resulted NEB replicas are used as an input to run Schoeck's and PAFI computations for entropy and free energy. Additionally, MD simulations were carried out to estimate the rate of the edge dislocation to overcome the interaction with the loop.

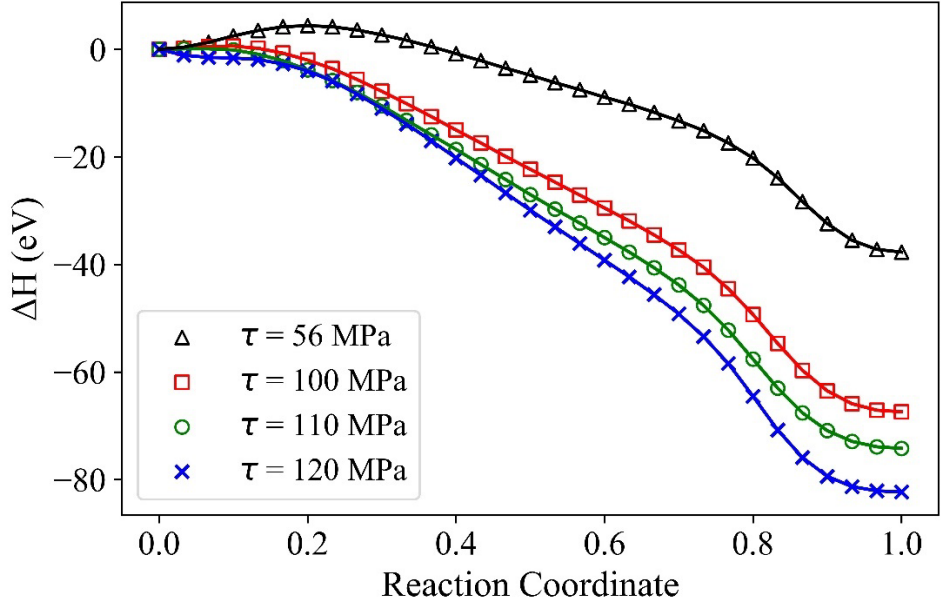

**Fig.2.** Enthalpy along the path at 56, 100, 110, and 120MPa resulting from NEB

*3.1. Entropy calculation with Schoeck's formalism*

The entropy profiles obtained using Schoeck's discretized formalism (Equation 11) are illustrated in Figure 3, which includes fitted splines to actual data points. It is evident that the majority of entropy values at temperatures lower than $0.3T_m$ of tungsten are negative. Considering differences in simulation setup, negative entropy is compatible with recent studies on the entropy of refractory alloys [20] for this range of temperatures. Moreover, as the shear stress increases, a combination of positive and negative entropy values emerges, leading to a more scattered distribution of data points. Notably, if the entropic term is comparable to the change in enthalpy, the fluctuation in entropy values will significantly influence the free energy barrier. Consequently, this effect could introduce a non-monotonic trend, which will be further discussed in section 3.2.

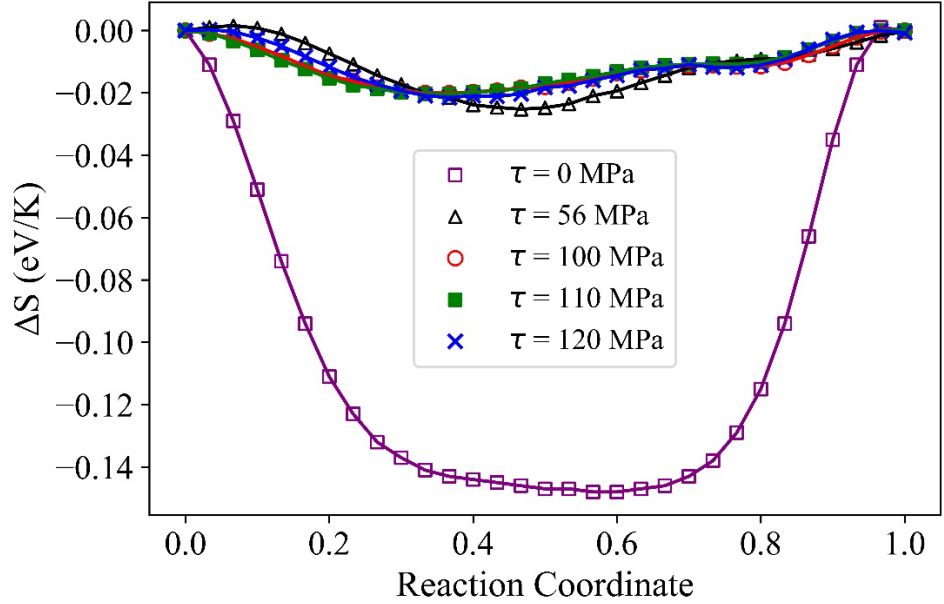

**Fig.3.** Entropy versus reaction coordinate in at 0, 56, 100, 110 and 120 MPa at 100 K.

*3.2. Free Energy Barrier Computation*

We computed the free energy along the MEP with Schoeck's formalism and compare it with PAFI at three different temperatures and two stresses, shown in Figure (4-a). Looking at the inset plot, PAFI shows relatively higher activation free energies compared to Schoeck's at each temperature at 120 MPa.

Figures (4-b to d) compare free energy barriers at T=180 and 240 K and 100 and 120 MPa shear stresses computed by Schoeck's formalism and PAFI. Comparing the two methods at 180 K and 100 MPa (Figure 4-b), Schoeck's predicted values result in a slightly higher barrier than PAFI, while the barrier at lower temperatures (30 and 60 K) is rather close for both methods, which is shown in Figure A.3 of the supplementary material.

Furthermore, Figures (4-c) and (d) present the free energy barrier for different temperatures at 120 MPa shear stress, slightly higher than the critical resolved shear stress value. We observe that the free energy barrier increases as the temperature increases from 180 to 240 K and the values for PAFI are slightly higher than Schoeck's for most replicas. The energy difference grows significantly as the reaction coordinate approaches the end. This phenomenon will be discussed later in the paper. Notably, as the temperature increases, a minimum of free energy develops between the initial state and the saddle point. Hence, the barrier must be measured accordingly, from the new minimum to the saddle point.

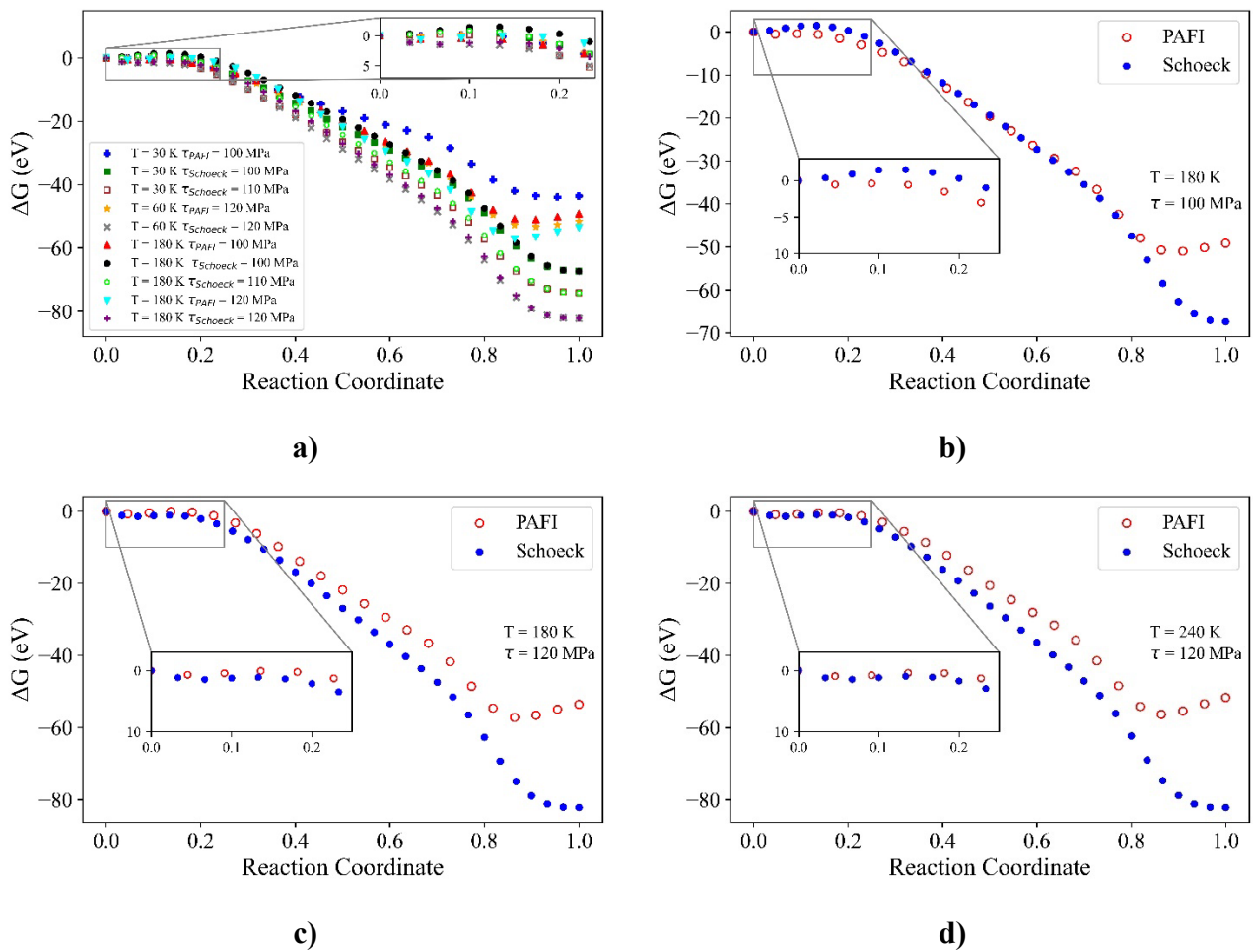

**Fig.4.** Free energy landscape computed by PAFI and Schoeck's methods for a) at 30 K, 60 K and 180 K, at 100, 110, and 120 MPa respectively. Comparing plot for the mentioned methods at constant shear stress of 100 MPa at b) 180 K. the same case near critical resolved shear stress at 120 MPa for temperatures of c) 180 K d) 240 K

As a general trend, we observe considerable consistency between Schoeck's and PAFI results throughout Figure 4.

Figure 5 specifically shows the activation free energy dependence with temperature at both 100 and 120 MPa. At 100 MPa, Schoeck's method demonstrates higher maximum values than PAFI's. At 120 MPa (Figure 5-b), the results from both methods are closer; the maximum values of PAFI are higher than Schoeck's, which is in contrast to Figure (5-a). Close to the $\tau_{CRSS}$ value, the activation free energy of PAFI fluctuates significantly more than the values obtained from Schoeck's approach. The trends, however, remain similar in both approaches. In Figure (5-c) we have compared the activation free energy for Schoeck's at 100, 110, and 120, for a broader temperature range (50 to 700 K), which exhibits stronger nonlinearity as shear stress is increased.

Based on Equation 11, we understand that the Gibbs free energy formalism should exhibit third-order dependence on temperature. Consequently, the optimal fitting curves for Schoeck's activation free energy under different stresses (100, 110, and 120 MPa) can be described as follows:

$$\Delta G_{a100_{Schoeck}} = -3.357 \times 10^{-9}T^3 - 3.461 \times 10^{-6}T^2 + 0.006082T + 0.5999 \qquad (12)$$

$$\Delta G_{a110_{Schoeck}} = -1.339 \times 10^{-9}T^3 - 5.304 \times 10^{-6}T^2 + 0.006278T + 0.1000 \qquad (13)$$

$$\Delta G_{a120_{Schoeck}} = -5.273 \times 10^{-8}T^3 + 2.7 \times 10^{-5}T^2 - 0.001445T - 0.005427 \qquad (14)$$

Comparing Equations 12 to 14, we understand that at 120 MPa the activation free energy expression has the lowest constant term which represents the barrier value at 0 K, i.e., the enthalpic barrier obtain with NEB, which is negative as the barrier is nullified above the $\tau_{CRSS}$. Following the same process, we can find the activation free energy expressions for PAFI keeping the same third order temperature dependence. The main reason behind this behavior will be discussed in section (3.3).

$$\Delta G_{a100_{PAFI}} = -1.906 \times 10^{-7}T^3 + 8.673 \times 10^{-5}T^2 - 0.009226T + 0.3402 \qquad (15)$$

$$\Delta G_{a120_{PAFI}} = -1.52 \times 10^{-7}T^3 + 5.68 \times 10^{-5}T^2 - 0.00565T + 0.144 \qquad (16)$$

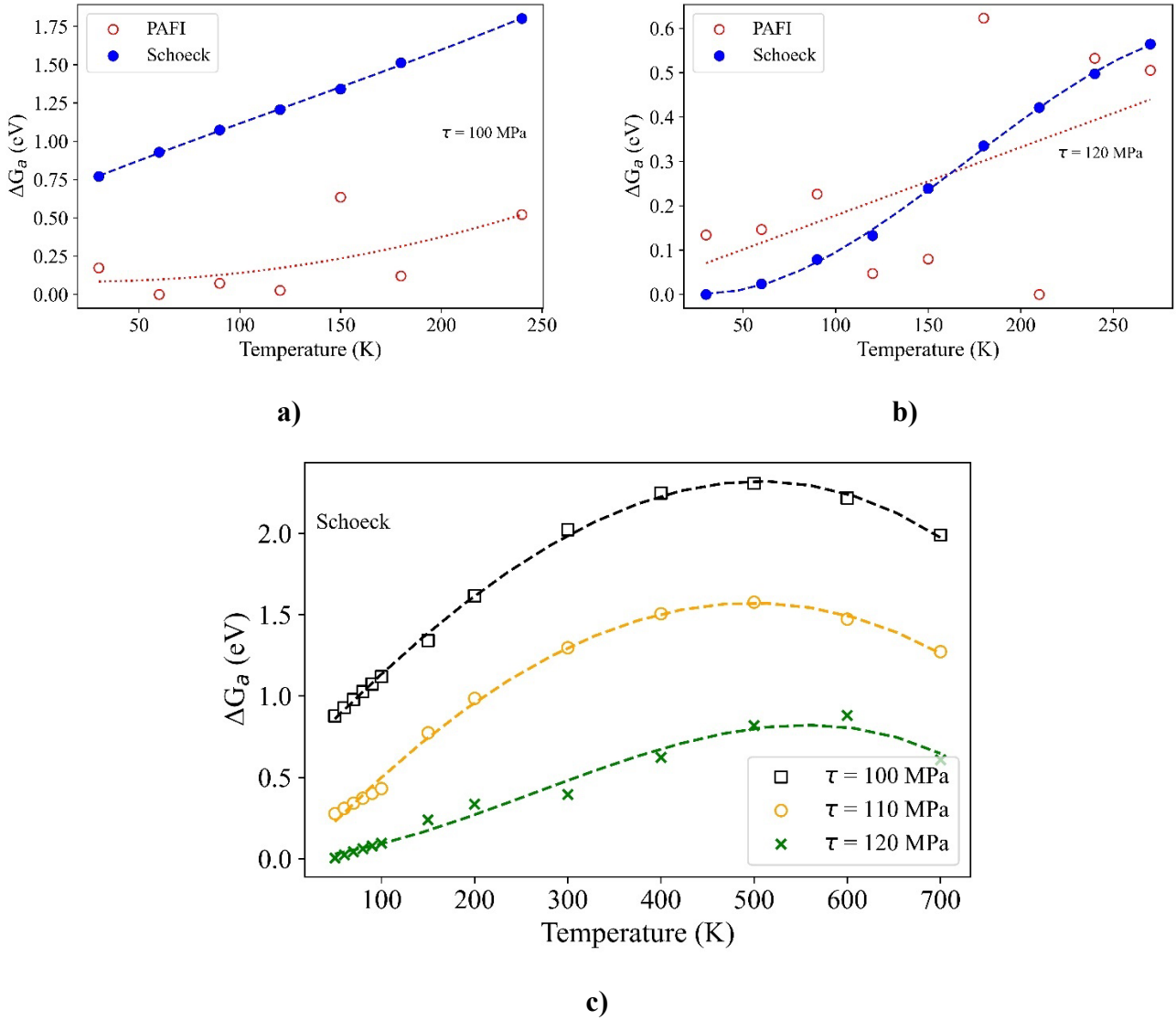

**Fig.5.** Activation free energy versus temperature resulting from **a)** 100, **b)** 120 MPa shear stress **c)** comparing plot of Schoeck's

*3.3. Rate of reaction calculation*

To compute the rate of reaction, in Schoeck's and PAFI methods, for the dislocation to overcome the barrier, we rely on transition state theory (TST) [40,41]:

$$\Gamma = \omega_0 \int_0^{L_d} \exp\left(-\frac{\Delta G_a^*(T,\tau)}{k_B T}\right) \frac{dl}{L_0} \qquad (17)$$

where $\omega_0$ is the pre-factor related to the attempt frequency, $L_0$ is the critical dislocation length for the segment to hop, $L_d$ is the total dislocation length equal to 184.521 (Å) in our simulation box, $\tau$ is the imposed shear stress, $\Delta G_a^*(T,\tau)$ is the activation free energy. The relation is fairly linear at low temperatures, presenting a maximum in all cases at temperatures between 400 and 600 K. For PAFI at 100 and 120 MPa the trend is non-monotonic and did not show any clear transition from linear to nonlinear behavior increasing stress (Figure 5-a and 5-b). In fact, it seems like the dependence of the activation free energy on temperature as given by Schoeck's formalism does not hold for PAFI. Moreover, based on Equation 2, we have determined that the temperature derivatives of elastic constants, which are influenced by the strain components, play a crucial role in defining the activation free energy. These factors collectively contribute to changes in the rate of reaction in response to variations in temperature at different stress levels.

In this study, a series of semi-analytical calculations, combined with the implementation of PAFI and MD simulations, were conducted in order to compute the rates for the dislocation to overcome the free energy barrier. The results are presented in Figure 6, showing the logarithm of the rate versus the inverse temperature. The reaction rate given by MD is quantified as the reciprocal of the average dislocation waiting times among the five independent simulations. The results of MD are presented for different shear stress levels, namely 100, 110, 120, and 125 MPa in the main and inset plots. When considering the cases of 100 and 110 MPa, both of which fall below the $\tau_{CRSS}$ threshold, it is observed that a close-to-linear trend emerges in the log-scale plot at elevated temperatures.

In the transition from 100 to 110 MPa, the magnitude of the slope of the best-fit line decreases from 0.27 to 0.13 eV. At 120 MPa and temperatures below 50 K, the negative slope changes to positive, with T = 50 K acting as the crossover point (inset plot in Figure 6). The positively sloped exponential line is referred to as a non-Arrhenius or anti-thermal relation. The black, orange, and green lines are for 100, 110, and 120 MPa shear stresses, respectively, obtained with Schoeck's approach. Schoeck's and PAFI show similar crossover points at 300 K and 200 K, respectively. We have used attempt frequencies of $1.62\times10^{30}$ s$^{-1}$, $2.1\times10^{27}$ s$^{-1}$, and $2.08\times10^{18}$ s$^{-1}$ for stress values of 100, 110 and 120 MPa for Schoeck's fitting to the MD values. Also, $1\times10^{25}$ s$^{-1}$ prefactor was defined for PAFI to match the crossover rate value at 120 MPa as given by MD. These prefactors are in the high range, as it has been observed in strained materials [19,42]. Figure 6 shows the non-monotonic temperature behavior of Schoeck's results at 120 MPa while monotonic for 100 and 110 MPa, which follows the MD simulation trend. PAFI results also display non-monotonic rates. Although, neither PAFI nor Schoeck's predict the rates obtained with MD, showing higher slopes, they capture the trend with crossovers from Arrhenius to anti-Arrhenius. However, there is a noticeable disparity of around 250 K in the crossover point compared to MD.

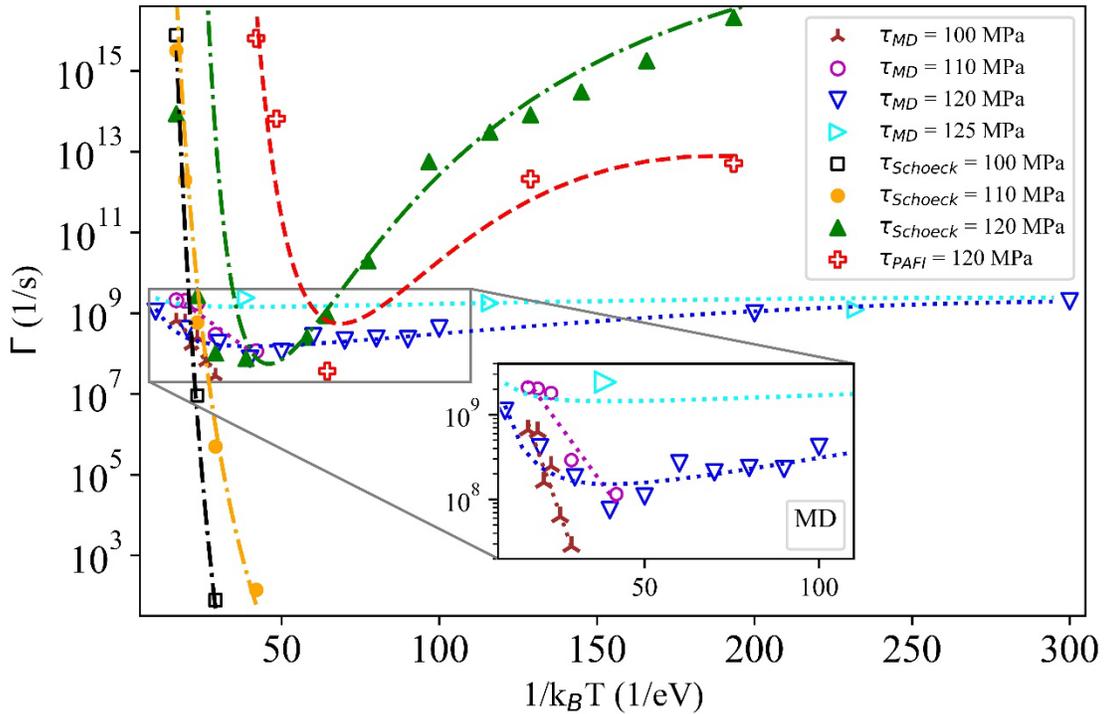

**Fig.6.** Rate of reaction versus temperature inverse fitted curves for atomistic simulation, PAFI and Schoeck's approaches in range of stress

Table 2 summarizes the results, highlighting the substantial impact of the shear stress on the predicted barrier value for dislocation crossing. It also shows the free energy expressions that are used to draw the dashed curves. An increase in shear stress leads to a decrease in the slope observed in the Arrhenius portion of the plots depicted in the figure, which is clearly shown for MD and Schoeck's following the actual rate values. This decrease indicates that the waiting time for dislocation crossing tends to decrease. Such a trend is more dominant around the $\tau_{CRSS}$ value (120 MPa).

Table 2. Fitted expressions rate plots for Atomistic Simulation, PAFI, and Schoeck

| | Shear Stress Value $\tau\ (MPa)$ | Rate fitting |
|---|---|---|
| Atomistic Simulation | 100 | $\Gamma = 7.06 \times 10^{10} \exp\left(-\dfrac{0.27}{k_B T}\right)$ |
| | 110 | $\Gamma = 2.11 \times 10^{10} \exp\left(-\dfrac{0.13}{k_B T}\right)$ |
| | 120 | $\Gamma = 10^{13} \exp\left(-\dfrac{3.68 \times 10^{-9} T^3 - 2.38 \times 10^{-6} T^2 - 0.0003T - 0.0033}{k_B T}\right)$ |
| | 125 | $\Gamma = 10^{13} \exp\left(-\dfrac{9.43 \times 10^{-10} T^3 - 5.59 \times 10^{-7} T^2 - 0.00066T - 0.000769}{k_B T}\right)$ |
| PAFI | 120 | $\Gamma = 1 \times 10^{25} \exp\left(-\dfrac{-1.52 \times 10^{-7} T^3 + 5.68 \times 10^{-5} T^2 - 0.00565T + 0.143691}{k_B T}\right)$ |
| Schoeck's | 100 | $\Gamma = 1.62 \times 10^{30} \exp\left(-\dfrac{-3.357 \times 10^{-9} T^3 - 3.461 \times 10^{-6} T^2 + 0.006082T + 0.5999}{k_B T}\right)$ |
| | 110 | $\Gamma = 2.1 \times 10^{27} \exp\left(-\dfrac{-1.339 \times 10^{-9} T^3 - 5.304 \times 10^{-6} T^2 + 0.006278T + 0.1000}{k_B T}\right)$ |

**Table 2**. Fitted expressions rate plots for Atomistic Simulation, PAFI, and Schoeck

| Shear Stress Value $\tau\ (MPa)$ | Rate fitting |
|---|---|
| 120 | $\Gamma = 2.08 \times 10^{18} \exp\left(-\dfrac{-5.273 \times 10^{-8}T^3 + 2.7 \times 10^{-5}T^2 - 0.001445T - 0.005427}{k_B T}\right)$ |

In Figure 7 the rate versus stress plot obtained from MD is shown. The activation volume can then be computed as $V_a = \dfrac{\partial \Delta G_a}{\partial \tau}\bigg|_T$. Basically, examining Equation 1, we can determine the activation volume as the stress derivative of the free energy expression. We can obtain $V_a$ from MD, where it corresponds to the slope of the curves in Figure 7. For instance, at 400 K and 110 MPa, this activation volume is calculated to be 1311 Å³, which is considerably smaller than the simulation cell volume. Furthermore, an slight decrease in activation volume is observed when the temperature increases, as depicted in Figure 7.

In Figure 8, we have plotted the activation entropy values against stress at various temperatures. This plot clearly illustrates the dependence of activation entropy on both stress and temperature. There is an increase in activation entropy by increasing stress from 100 to 120 MPa. The linear fits for each temperature have been included to illustrate the increasing trend with respect to stress values.

Analyzing the breakdown of the free energy as explained in Section 1, it became evident that four distinct terms - potential, elastic, plastic, and entropic - play a role. To assess the significance of each term, Figure (9-a) presents the energy distribution along the path for the latter three terms, while Figure (9-b) illustrates the distribution of potential energy at various stress levels along the NEB coordinate. From a theoretical standpoint, Figure (9-a) demonstrates that the most significant contributing term to the activation free energy is the plastic energy shown in red, yellow, and green, in contrast to the elastic and entropic terms which are relatively small. In Figure (9-b), the term $\Delta E_0$ is shown, as obtained from NEB with no applied stress and with the values for 100, 110, and 120 MPa adding the elastic and plastic work components. The discrepancy comes from the uncertainty in computing the elastic and plastic terms.[30]

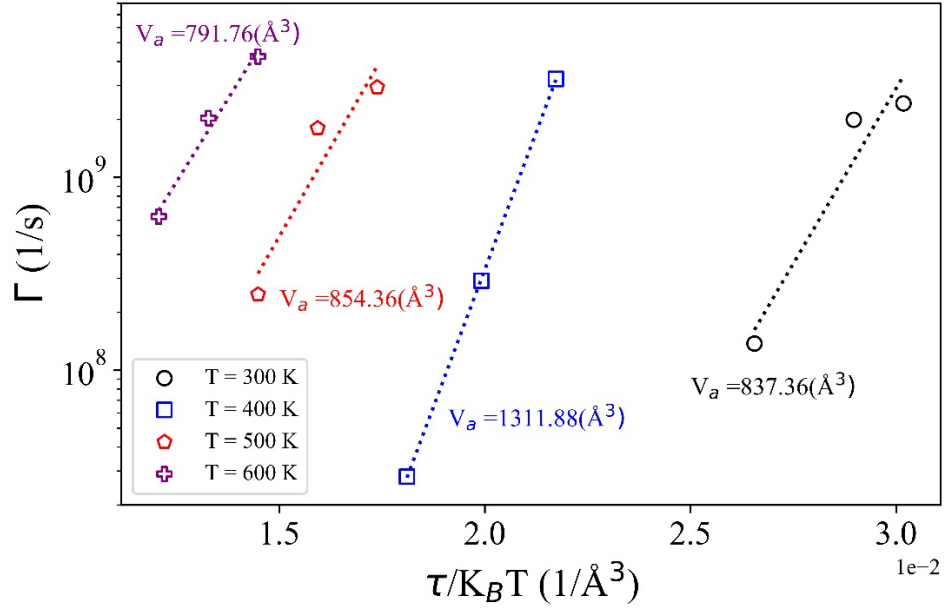

**Fig.7.** Rate versus stress from 300 to 600 K obtained from MD simulations. Dashed lines are linear fits as a guide to the eyes with an estimate of the average slope related to $V_a$.

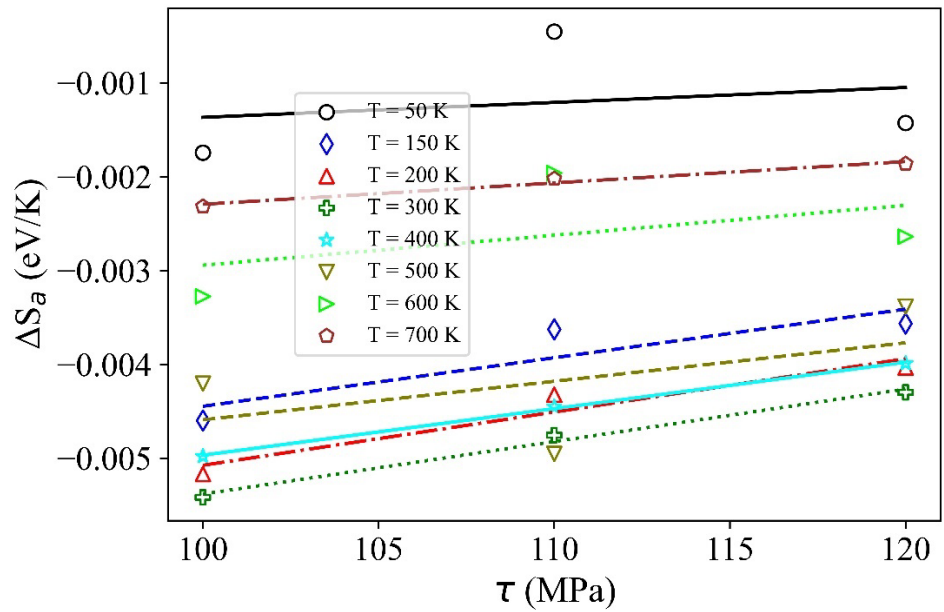

**Fig.8.** Activation entropy vs stress for temperatures from 50 to 700 K in Schoeck's method

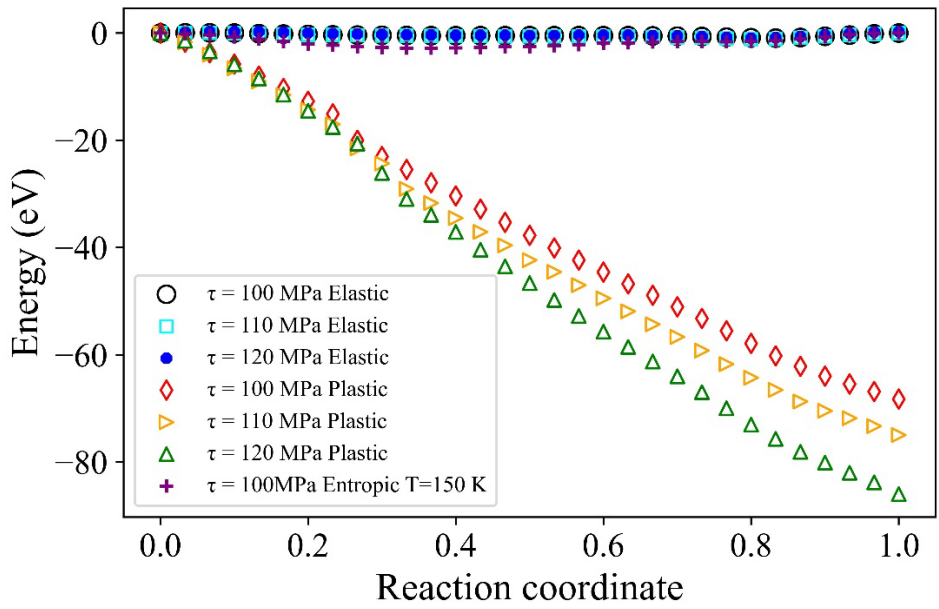

a)

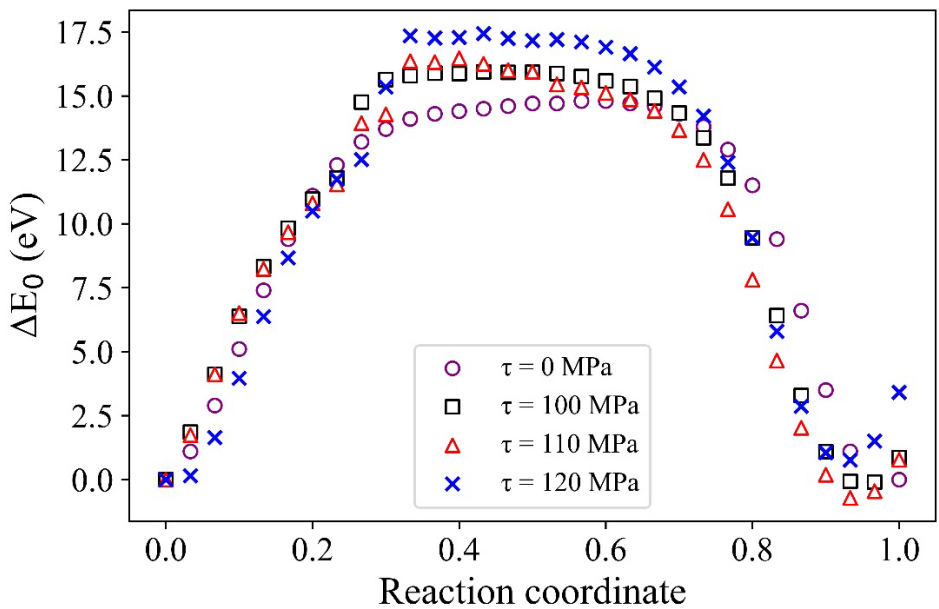

b)

**Fig.9. a)** Energy components along the MEP resulted from theoretical computations compared to entropic energy at 150 K and 100 MPa predicted **b)** Energy change associated with the displacement of a dislocation line adjacent to a prismatic loop.

In Figure 10-a, we compare the contributions of each energy term to the activation volume prediction based on theoretical calculations. As mentioned earlier, in Figure 7, the activation volume is the stress derivative of the activation free energy at constant temperature [17], $V_a = \frac{\partial \Delta G_a}{\partial \tau}\Big|_T = \int_V [\tilde{\epsilon}^s - \tilde{\epsilon}^i] dV + (\bar{b} \otimes \bar{d})l - T\frac{\partial \Delta S_a}{\partial \tau}$. As shown above, the activation entropy values depend on stress. The two entropic and plastic terms vary with temperature, generally exhibiting an increasing trend up to 500 K, followed by a slight decrease. Also Figure 10-b shows the activation volume change with respect to stress.

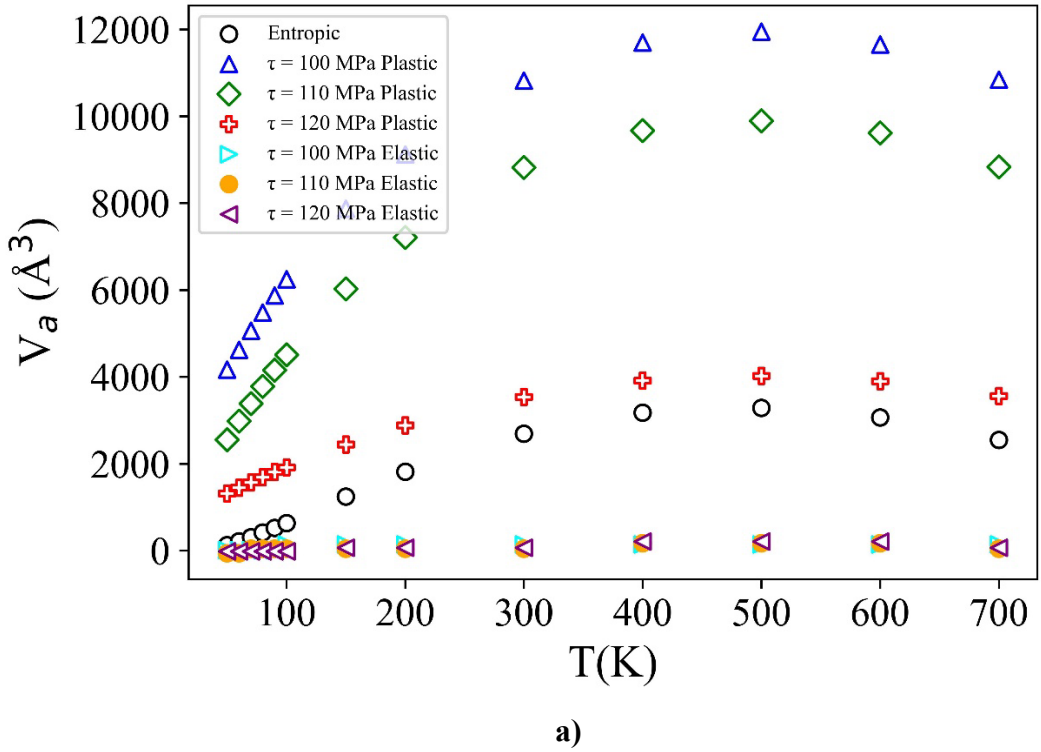

a)

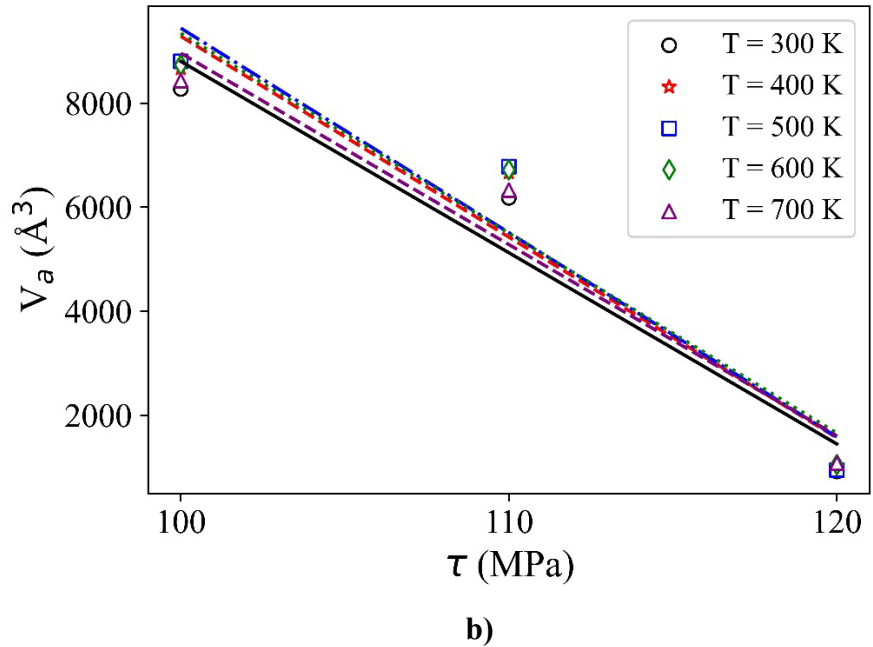

**Fig.10.** a) contribution of each terms in activation free energy expression (Equation 1) in activation volume value b) activation volume values versus stress values for different temperatures

Figure 11 illustrates the computed activation entropy values corresponding to PAFI and Schoeck's approaches at 120 MPa versus temperature. PAFI's values do not follow a clear trend, unlike Schoeck's. As indicated by Equation 11, we anticipate a second-order dependence of the activation entropy on temperature for each stress level, as shown in Fig. 11 with lines being best fits to the data (equations 18 to 20) at 100, 110, and 120 MPa, respectively. In the range of temperatures studied, $\Delta S_a$ is not monotonic, presenting a minimum at around 300 K, which coincides with the crossover temperature between the Arrhenius and anti-Arrhenius behavior of the rates.

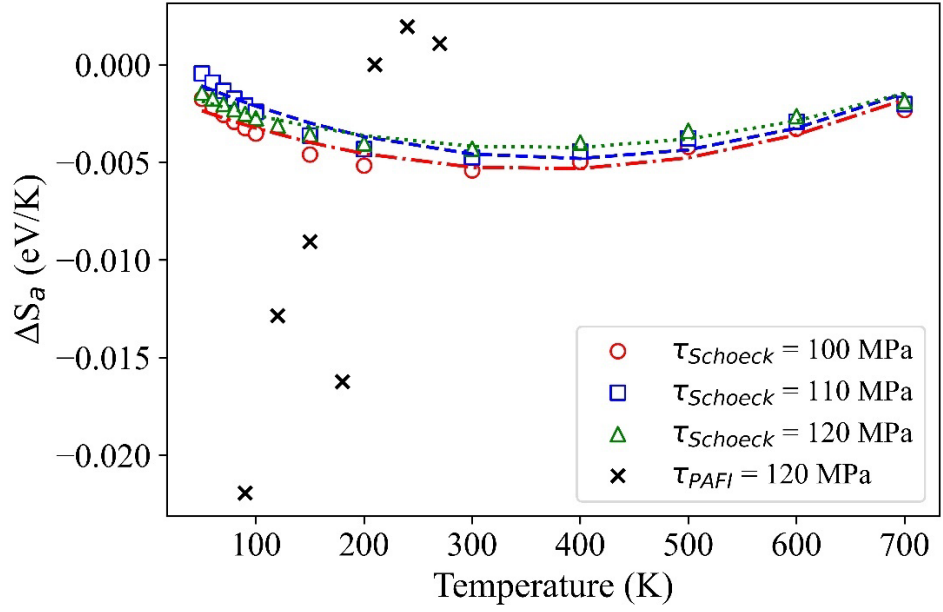

**Fig.11** Activation entropy values for different temperatures predicted by Schoeck's and PAFI methods at 100, 110, and 120 MPa

$$\Delta S_{a100_{Schoeck}} = 3.113 \times 10^{-08}\, T^2 - 2.24 \times 10^{-05} T - 0.001311 \tag{18}$$

$$\Delta S_{a110_{Schoeck}} = 3.343 \times 10^{-08} T^2 - 2.566 \times 10^{-05} T - 0.0001108 \tag{19}$$

$$\Delta S_{a120_{Schoeck}} = 2.466 \times 10^{-08} T^2 - 1.788 \times 10^{-05} T - 0.001037 \tag{20}$$

## 4. Discussion

In this paper, we have investigated the effects of stress and temperature on the free energy barrier, activation entropy, activation volume, and rate of reaction of an edge dislocation overcoming the far field interaction with a prismatic loop in BCC tungsten. We compared the results from MD, PAFI, and theoretical Schoeck's formalism. These methods were applied across various stress values around the critical resolved shear stress of BCC tungsten to visualize how the mentioned properties change with varying temperatures.

Both Schoeck's and PAFI predict similar trends for the free energy landscape between initial and saddle point states for the shear stresses tested in this study. The free energy landscape computed by Schoeck's closely resembles the predictions made by PAFI at 100 and 120 MPa along the minimum energy path, encompassing a total of 31 replicas, although PAFI data point exhibit noisier trends than Schoeck. As we approach the end of the reaction coordinate, PAFI exhibits higher values than Schoeck's. The energy difference between the initial and final states primarily stems from changes in plastic energy, reflected in the MEP at different stresses. Consequently, this discrepancy is likely due to the deviation of PAFI from the MEP obtained with NEB. Both methods use an initial MEP to compute free energies, with the major difference being in how the free energy is computed. Schoeck's requires the calculation of the entropy depending on temperature, elastic constants as a function of temperature, strain components, and atomic volume. In terms of computation cost, Schoeck is more time-effective than PAFI. The atomic strains were computed by Ovito along with the atomic volumes. However, PAFI requires the estimation of the partition function for each replica running dynamics, which becomes more computationally expensive. Somehow, PAFI also deviates from the expected Arrhenius trend at lower stress values. We demonstrated in Figure 2 that as the stress values increase from 56 to 120 MPa, the barrier height decreases. Furthermore, the computed entropy values along the path exhibit an increase as the shear stress is raised (see Figure 8). Within the Schoeck's formalism, the components of the elastic constant tensor have been shown to exhibit strain dependence, along with sensitivity to temperature, both of which notably impact the entropy values.

The rate of reaction analysis from MD results reveals that in the proximity of $\tau_{CRSS}$, we observe a change in the trend from anti-Arrhenius to Arrhenius behavior as the temperature increases. This shift is attributed to the very small or null enthalpy barrier that the dislocation needs to overcome. As a result, the entropic term plays a significant role in defining the mechanism of dislocation motion. The temperature at which the crossover from anti-Arrhenius to Arrhenius occurs is approximately 50 K in MD. In Schoeck's and PAFI rate plots, a similar behavior is observed, but at higher crossover temperatures around 300 K and 200 K, respectively. In PAFI, we observe that the results for the rate near $\tau_{CRSS}$, specifically at 120 MPa, exhibit more variability compared to Schoeck's still showing the crossover in trends. Furthermore, according to Schoeck's approach, the rate is not Arrhenius even for stresses significantly lower than $\tau_{CRSS}$ but the crossover appears in an extremely low-rate regime (see Fig. A.4 supplementary material).

Another notable difference among the methods was the barrier height, which is higher in Schoeck's and PAFI compared to MD. This leads to a steeper slope of the rate plots. While theoretical computations indicate a higher expected temperature for the minimum rate, the computational cost is significantly lower than that of MD. Additionally, this method can provide separate insights into entropy, enthalpy, elastic, and plastic energy contributions in the deformation mechanism. It is worth noting that MD was utilized to obtain temperature and stress values for determining the dislocation waiting time across the barrier, yet it could not display individual components as calculated using the Schoeck's formalism. The assessment of activation volume was conducted using both MD and Schoeck's methods, as depicted in Figures 7 and 10. Although PAFI could compute free energy and entropy values, it fails to correctly compute activation volume since it shows noisy activation free energy profiles in most of the stress values shown in Figure (5-a) and (5-b). The MD results indicate that higher temperatures lead to a reduction in the activation volume. However, a more comprehensive insight into the results, accounting for each energy component contribution through Schoeck's approach, reveals that the plastic and entropic terms play a pivotal role in determining the activation volume, while the contribution of elastic terms appears to be negligible.

A potential explanation for the difference in rates between Schoeck's, PAFI, and MD is the difference in the path followed by the system to overcome the energy barrier. The MEP reported here was obtained starting from a linear interpolation between initial and final configurations. The energy landscape with dislocations is potentially rough and might lead to significant deviations from the linearly interpolated guess. As mentioned in section 2, we also tried as initial input for the NEB snapshots from the MD trajectory, but the results did not vary significantly. Still, an accurate MEP seems crucial for the prediction of the transition rates, although in configurations involving dislocations the task is far from trivial. Besides, the rate of reaction that PAFI and Schoeck's compute is based on the transition state theory which accounts for thermodynamic properties, while MD computes the rate based on the observation of the waiting time from a dynamic trajectory at different temperatures and stresses. Future work will analyze the detail role of the path in the estimation of the free energy and the calculation of the rates.

Schoeck's formalism has deep consequences on our ability to model ensemble of dislocations at the mesoscale. Current methodologies are based on elastic energies to study the evolution of an ensemble of dislocations. However, following Schoeck's theory, materials with a strong dependence of its elastic constants on temperature and with strong thermal expansion might show an important entropic component that will modify the free energy landscape and therefore the driving force. The analysis of the role of entropy on mesoscale models and how it compares with the enthalpic terms will also be the subject of future work.

As a last remark, our analysis seems to imply that between the purely thermally activated regime at low temperatures and applied stresses and the phonon drag regime at high temperatures and/or higher stresses, there is a regime dominated by the activation entropy. How important this intermediate regime is will be material dependent, related to the dependence of the elastic constants with temperature (and potentially strain), and could significantly alter the behavior of dislocations overcoming barriers induced by long-range interactions.

## 5. Summary and concluding remarks

In this study, we have presented a comprehensive investigation into the behavior of an edge dislocation overcoming the interaction with an $a\langle 100 \rangle$ prismatic loop in pure tungsten under specific temperature conditions and line defect configurations at different stress levels, specifically near the critical resolved shear stress. Our analysis involves a combination of theoretical calculations following Schoeck's formalism to obtain entropy changes induced by internal strains, molecular dynamics simulations, and Projected Average Force Integrator (PAFI) simulations. We computed the activation enthalpy using the nudged-elastic band method for different applied stresses and the activation entropy following Schoeck's approach. Relying on transition state theory (TST) we computed the rates for the dislocation to overcome the far field interaction with a prismatic loop and compared the results with MD. We also used PAFI to predict the activation free energy and TST to predict the rate for the dislocation-obstacle bypass. We checked the temperature and shear stress effects on the activation entropy, activation enthalpy, activation volume, and activation free energy. We observe that the activation enthalpy and free energy decrease as the shear stress increases, while its magnitude exhibits a third order dependence on temperature according to the theoretical approach. Also, the contribution of the activation entropy on the activation volume was found to be highly temperature dependent while the elastic and plastic energies almost did not change with temperature.

Schoeck's formalism and PAFI results agree remarkably well at 120 MPa and were able to replicate the observed trends in the rate obtained with MD, while PAFI's noisy trend in activation free energy leads to large uncertainty in the rate prediction at other stress values. In addition, for the purpose of comparing the three different approaches, as presented in Figure 6, we highlight that while all three methods successfully capture the Arrhenius/non-Arrhenius transition near the critical resolved shear stress, PAFI and Schoeck's methods show different crossover temperatures and barrier values compared to MD. A deeper analysis of the transition path or other causes of lower rates (such as re-crossings) is needed to fully understand this discrepancy. Also, assessing the performance of these three methods, we understand that Schoeck's method exhibited lower computational time compared to PAFI and MD to obtain changes in entropy, the rate of reaction, and activation volume.

## 6. Acknowledgments

M.N., S.S. and E.M. thank support from the startup package at Clemson University. Clemson University is acknowledged for generous allotment of compute time on the Palmetto cluster. Creative Inquiry program at Clemson University is also acknowledged. E.M. (partial support), S.B., and D.P., acknowledge support from the U.S. Department of Energy, Office of Science, Office of Fusion Energy Sciences, Office of Advanced Scientific Computing Research through the Scientific Discovery through Advanced Computing (SciDAC) project on Plasma Surface Interactions under Award No. DE-SC0008875. The authors thank Nithin Mathew for many useful discussions.

**Data availability**

All data and codes used in the simulations performed in this paper are available from the corresponding author upon reasonable request.

**Supplementary Material**

Strain distribution:

In Figure A-1 the strain tensor components, resulted from 120MPa shear stress, has been shown versus atomic ID which is important for curve-fitting and selection of the proper relation.

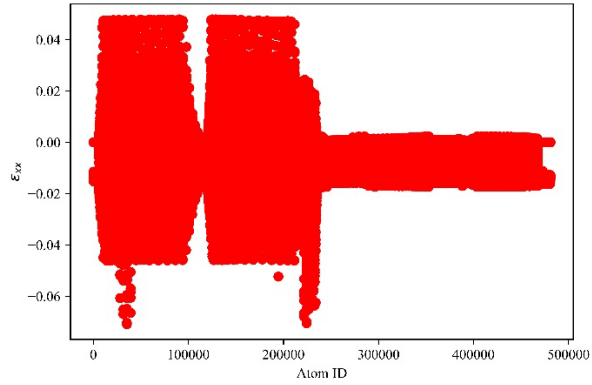

(a)

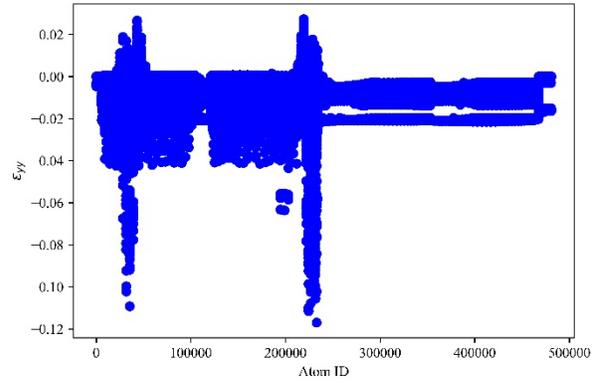

(b)

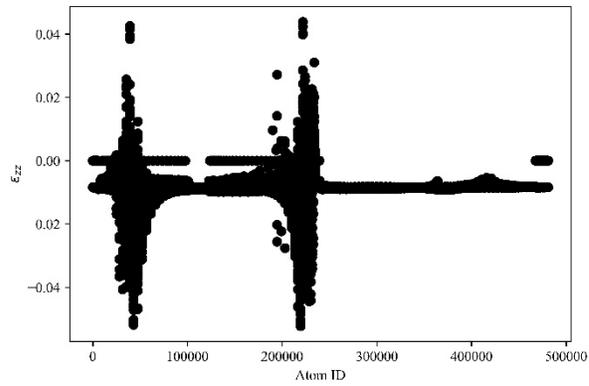

(c)

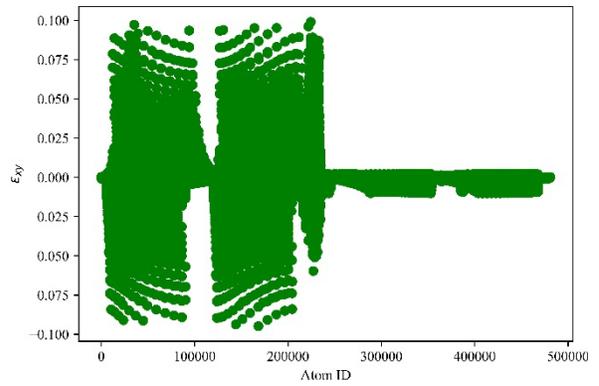

(d)

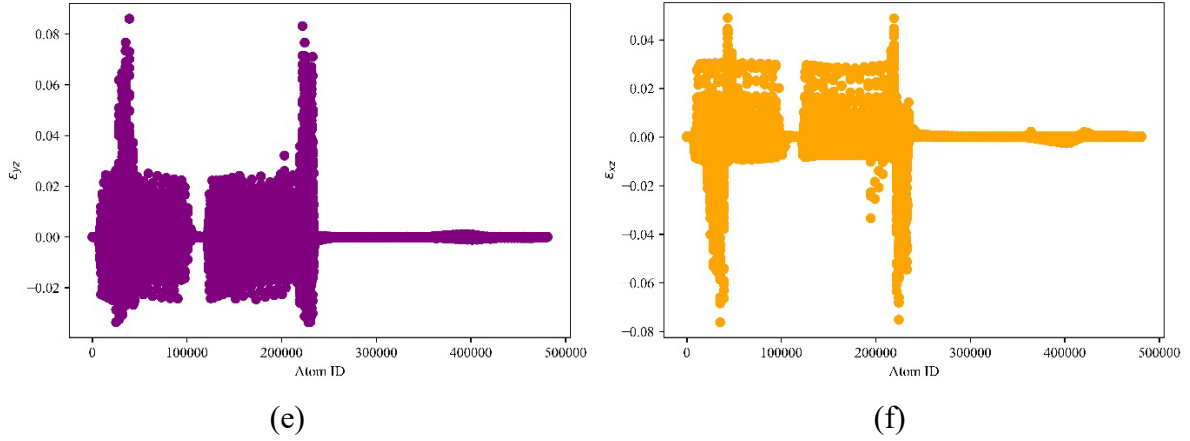

(e)                                  (f)

Fig. (A.1). Strain components for different atoms in the simulation cell

Stress strain curves for actual data of different components are shown in Figure (A.2). The fitted curves include extrapolation beyond strain value of 0.039 in order to capture corresponding elastic constant expressions which are in fact strain derivatives of stress or tangent to curves at each point (Equations 3-9).

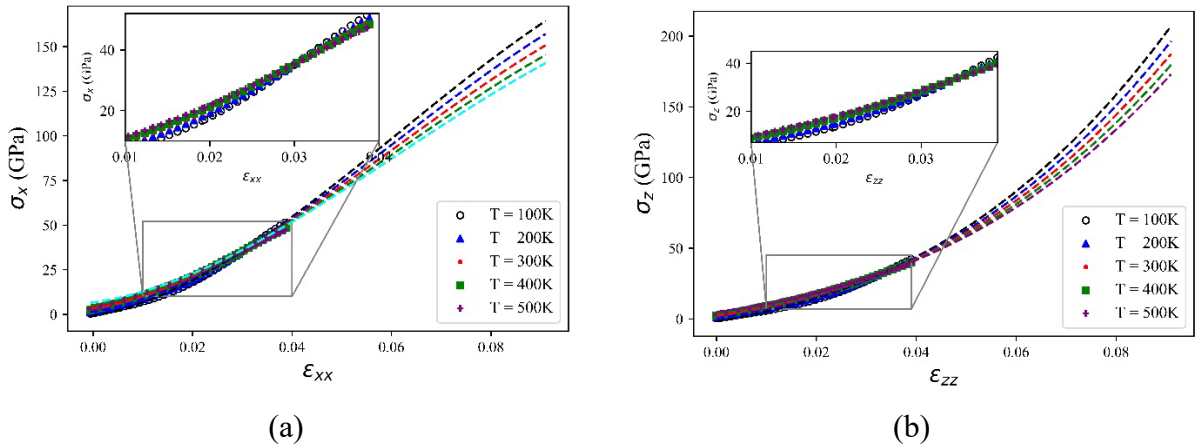

(a)                                  (b)

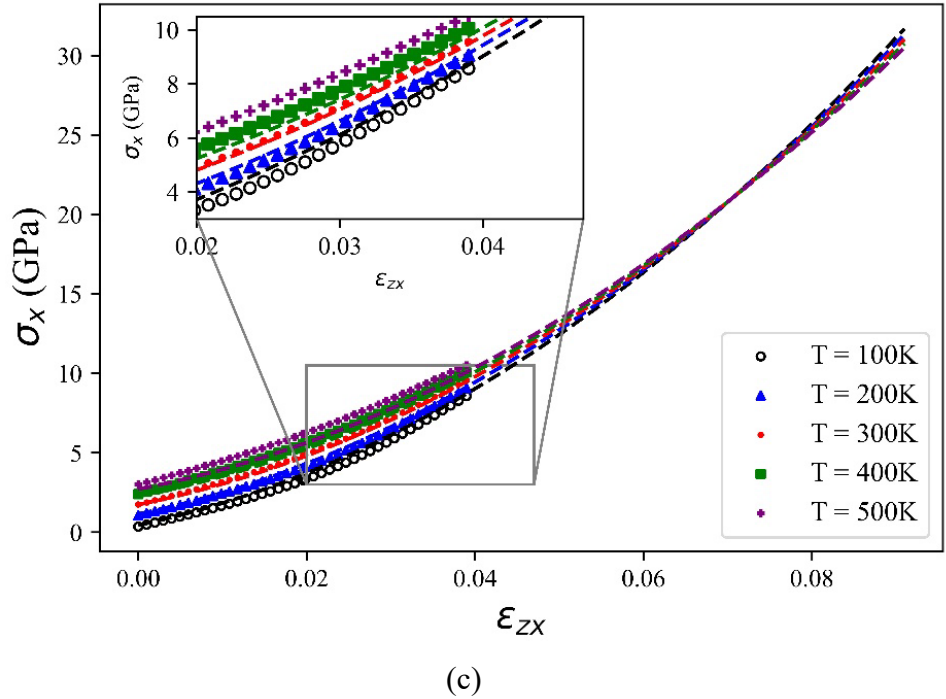

(c)

Fig. (A.2). Stress vs strain curves for different component

Figure A.3 shows the free energy landscape computed by PAFI and Schoeck's formalism at low temperatures.

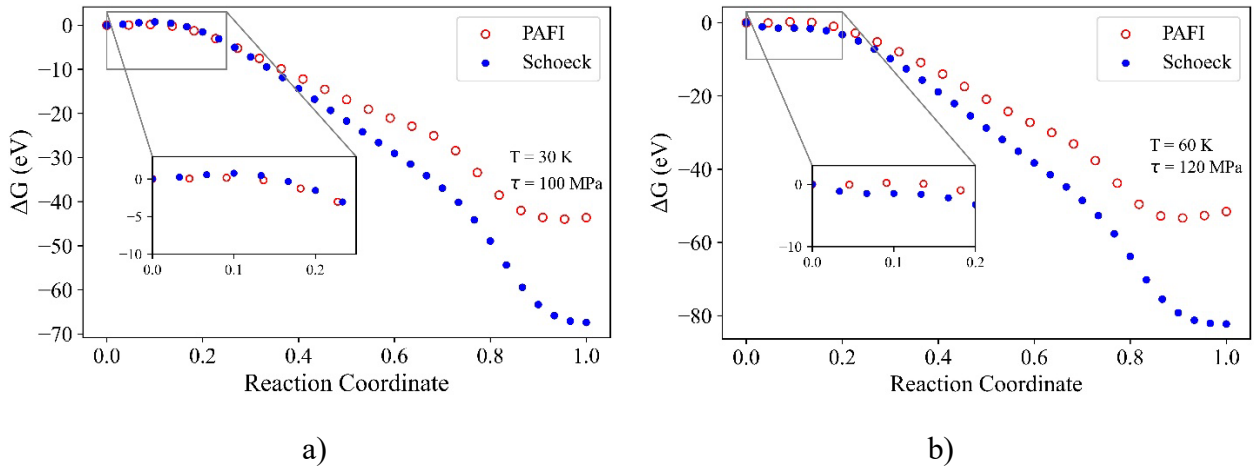

a)                                                    b)

Fig. (A.3). Free energy landscape computed by PAFI and Schoeck's methods

Figure A.4 shows the Schoeck predicted rate of reaction for the temperature range from 50 to 700 K.

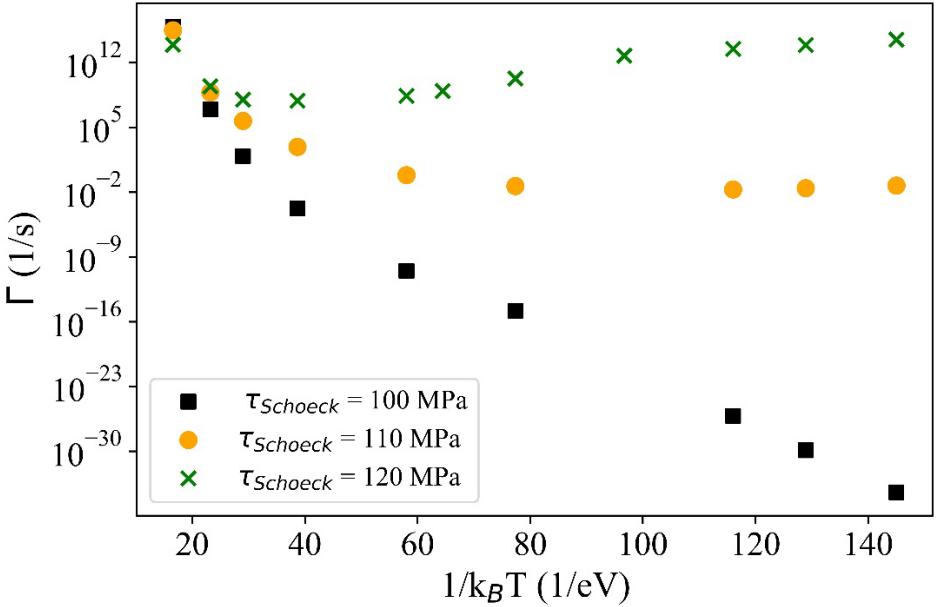

Fig. (A.4). Rate of reaction versus inverse temperature for Schoeck approach